\documentstyle[sprocl,epsf]{article}

\def\be{\begin{equation}}
\def\ee{\end{equation}}
\def\bea{\begin{eqnarray}}
\def\eea{\end{eqnarray}}
\newcommand{\Mc}{{\cal M}}
\newcommand{\Ms}{M_{\odot}}

\newcommand{\hf}{\tilde h}
\newcommand{\rf}{\tilde r}

\def\ltsima{$\; \buildrel < \over \sim \;$}
\def\simlt{\lower.5ex\hbox{\ltsima}}
\def\gtsima{$\; \buildrel > \over \sim \;$}
\def\simgt{\lower.5ex\hbox{\gtsima}}

\begin{document}

\title{COALESCING BINARIES AND DOPPLER EXPERIMENTS}

\author{A. Vecchio}

\address{
Max Planck Institut f\"{u}r Gravitationsphysik,
Albert-Einstein-Institut\\
14473 Potsdam, Germany}

\author{B. Bertotti}

\address{
Dipartimento di Fisica Nucleare e Teorica, Universit\`a di Pavia \\
27100 Pavia, Italy}

\author{L. Iess}
\address{Dipartimento di Ingegneria Aerospaziale, Universit\`a
"La Sapienza" \\ 
00158 Roma, Italy}

\maketitle\abstracts{We discuss the sensitivity of the CASSINI 
experiments to gravitational waves emitted by the in-spiral of compact
binaries. We show that the maximum distance reachable by the instrument
is $\sim 100$ Mpc. In particular, CASSINI can detect massive black hole binaries
with chirp mass $\simgt 10^6\,\Ms$ in the Virgo Cluster with signal-to-noise
ratio between 5 and 30 and possible compact objects of mass $\simgt 30\,\Ms$
orbiting the massive black hole that our Galactic Centre is likely to harbour.
}

The Doppler tracking \cite{EW} of interplanetary spacecraft in the only technique
presently available to search for gravitational waves (GW) in the low frequency
regime. During the past decade,
a number of experiments, lasting from some day to a month, have been carried 
out, the most recent involving the spacecraft GALILEO, MARS-OBSERVER and 
ULYSSES \cite{pap1,BBgr14}. The CASSINI probe,
due to be launched on October 6th 1997 and equipped with an improved radio link
(in X and K$_a$ band, $8.4$ and $34\,{\rm GHz}$, respectively), will 
perform three 40-days data acquisition runs between 2000 and
2004. Here we discuss the expected sensitivity of these experiments
to GWs emitted during the in-spiral of binaries of compact objects.

In a Doppler experiment a GW of amplitude $h$ produces at the detector output 
a signal $\tilde s(f)$, that, in the convenient Fourier domain, reads \cite{EW}
\be
\tilde s(f) = \rf_{\theta}(f;T) \hf(f)\,,
\label{sf}
\ee
where $\tilde r_{\theta}(f;T)$ is the characteristic {\it three-pulses
response}, which depends on $T$, the round-trip light-time of the radio link
out to the distance 
of the probe, and on $\theta$, the angle between the spacecraft and the
source. Focusing attention on the radiation
emitted during the in-spiral of a binary toward its final
coalescence, $\tilde h$ reads \cite{BVI}
\be
\hf(f) = \left(\frac{5}{96\pi^{4/3}}\right)^{1/2}\,
Q\,\frac{\Mc^{5/6}}{D}\, f^{-7/6} e^{i \Psi(f)}\,,
\label{hf}
\ee
where $\Mc$ is the chirp mass of the binary, $D$ the distance and $Q$ a 
function of the orientation of the source and the GW polarization; 
$\Psi(f)$ is the GW phase.
Eq. (\ref{hf}) is valid up to some frequency $f_{\rm isco}$,
after which the final coalescence sets off with emission of a
burst, whose structure is still quite unknown; indeed, we
cut-off the signal (somehow arbitrarily) at $f = f_{\rm isco}$.

The signal-to-noise ratio (SNR) $\rho$ measured at the output of 
the optimal matched filter reads:
\be
\rho^2 = 4\,\int_{f_b}^{f_e}\,\frac{|{\tilde s}(f)|^2}{S_n(f)}\,df
= 
\left(\frac{5Q^2}{24\pi^{4/3}}\right)\,\frac{\Mc^{5/3}}{D^2}
\,\int_{f_b}^{f_e}\,\frac{f^{-7/3}\,|\rf_{\theta}(f)|^2}{S_n(f)}\,df\,,
\label{snropt} 
\ee
where $S_n(f)$ is the noise spectral density of the detector;
$f_b$ and $f_e = \min[f_b\,(1-T_1/t_n)^{-3/8},\,f_{\rm isco}]$ are the instantaneous 
frequencies of $h(t)$ at the
the beginning ($t = 0$, for convention) and at the end 
($t = T_1$) of the data set; here $T_1$ is the time of observation and 
$t_n$ the time to coalescence from $f_b$ (in general $t_n\gg T_1$). 
In the ideal case in which we neglect the finite bandwidth of
the signal and of the instrument ($f_b\rightarrow 0$, $f_e\rightarrow \infty$)
and the different frequency response due
to the structure of the noise and of the three-pulses filter
($S_n =$ const., $\theta = \pi/2$), we can derive and {\it ideal} SNR, $\rho_{\rm id}$,
that depends only on fundamental parameters of the source and the detector;
for CASSINI typical values it reads
\be
\rho_{\rm id} \simeq 3 
\left(\frac{D}{100 \,{\rm Mpc}}\right)^{-1}\,
\left(\frac{\Mc}{10^7\Ms}\right)^{5/6}
\left(\frac{T}{10^4 {\rm sec}}\right)^{2/3}
\left(\frac{S_n}{9\times 10^{-26}\,{\rm Hz}^{-1}}\right)^{-1/2}\,.
\label{rhoid}
\ee
This result, although obtained in a simplified (and optimistic) case, clearly shows 
that the instrument is sensitive to binaries out to the distance of the 
Virgo Cluster ($D\simeq 17\,{\rm Mpc}$) and beyond.
Of course the reference value (\ref{rhoid}) is not attained in a real
experiment, due to the limited bandwidth $(f_b,f_e)$, the strong oscillations 
of $\rf_{\theta}(f;T)$, that, 
depending on $\theta$ and $(f_b,f_e)$ can suppress the signal at the detector output, 
and the increase of $S_n(f)$ at low frequencies. We can 
rigorously describe the degradation
of SNR, with respect to $\rho_{\rm id}$, introducing the
{\it detector efficiency function} 
\be
\Upsilon(f_bT,f_eT,\theta) \equiv \frac{\rho}{\rho_{\rm id}}\le 1\,.
\ee
Its main features are the following \cite{BVI}: (i) $\Upsilon$ is peaked 
at $f\sim 1/T$; as a consequence,
for $f_b$ larger then a few $1/T$, $\Upsilon$ is severely depressed;
searches for signals in past Doppler data \cite{BBgr14} have always concentrate
on the band $f > 1/T$, therefore missing the most sensitive portion
of the observable spectrum; (ii) only broad band searches  
(say $f_e \simgt 2\,f_b$) of chirping signals
can produce $\Upsilon \rightarrow 1$; in order to reach distant 
sources, one can confine the analysis to binaries with mass
$\Mc \simgt 3.3\times 10^7\,
(f_b/10^{-5}\,{\rm Hz})^{-8/5}\,(T_1/40\,{\rm days})^{-3/5}\,\Ms$;
(iii) the optimal direction of observation $\theta$, for given $\Mc$ and $D$,
depends on the frequency band $(f_b,f_e)$ swept by the signal.

\begin{figure}[t]
\centerline{\mbox{\epsfysize=10.cm
\epsffile{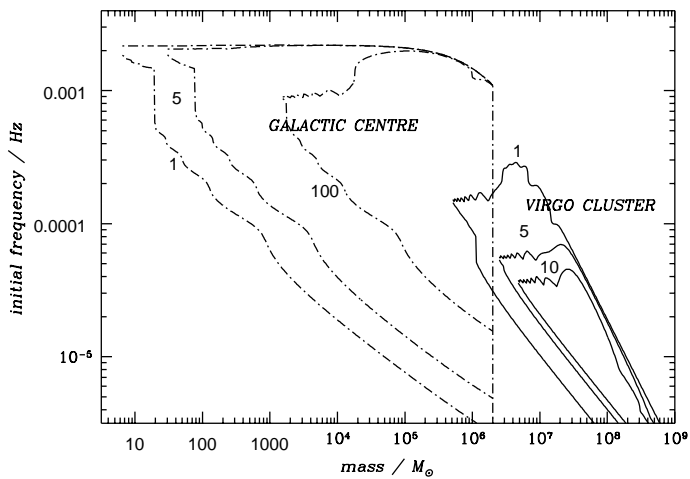}}}
\caption{\label{fig:sens}
Detectable binaries in the Virgo Cluster 
(solid line; contour plots of SNR $= 1\,,5\,,10$ in the plane $(\Mc,f_b)$; 
$m_1 = m_2$, $\cos\theta \simeq -0.04$, $D = 17\,{\rm Mpc}$)
and our Galactic Centre (dot-dash line;
contour plots of SNR $= 1\,,5\,,100$ in the 
plane $(m_2,f_b)$ for $m_1 = 2\times 10^6\,\Ms$;
$\cos\theta \simeq -0.85$, $D = 8\,{\rm kpc}$) during CASSINI second
experiment ($T\simeq 6796\,{\rm s}$, $T_1 = 40\,{\rm days}$).
}
\end{figure}

We apply now the previous results to the CASSINI experiments for 
two specific astrophysical 
targets: the Virgo Cluster and our Galactic Centre.  
In both cases the sensitivity of the three data sets turns out to be quite
comparable and we discuss only the second one.
The results are summarized in Fig. \ref{fig:sens}.
%
%
%
%

For the Virgo Cluster, binaries with $\Mc \simgt 2\times 10^6\,\Ms$ and
$f_b \simlt 7\times 10^{-5} \, {\rm Hz}$ can be detected at SNR $\ge 5$,
and in the region of the plane $(\Mc,f_b)$ around $(10^{-5}\,{\rm Hz},
5\times 10^7\,\Ms)$ signals produce ${\rm SNR} \simeq 30$. These
results hold for systems of comparable mass; decreasing the mass ratio 
$m_2/m_1$, the sensitivity region in the plane $(\Mc,f_b)$ shrinks and 
for $m_2/m_1 < 0.01$ 
the instrument is not able to detect any source (at $D = 17\,{\rm Mpc}$).
The detector can therefore reach, at ${\rm SNR} = 5$, a distance 
$\simeq 100\,{\rm Mpc}$ for $m_1 \sim m_2 \sim 10^8\Ms$. According
to current estimates of the event rate of super-massive black hole binary
coalescence \cite{Lisavecch}, we can conclude that even with this new generation
of Doppler instruments the chance of detection is probably low;
however 
(i) when accurate templates for the signals emitted during the final
black hole merger will be available, the search depth can
considerably increase; (ii) the results reported in Fig. \ref{fig:sens}
have been obtained using the conservative estimate $S_n\propto f^{-2}$ for
$f < 10^{-4}\,{\rm Hz}$; improvements in the data
pre-processing could lead to a flatter noise curve at 
very low frequencies, enhancing the sensitivity in that very
crucial band; (iii) $T_1$ could be extended.

We consider now galactic observations. Based on evidences
of a massive black hole in our Galactic Centre \cite{EG}, we investigate
the ability of CASSINI of picking up signals emitted by a compact object
of mass $m_2$ orbiting the central black hole $m_1 = 2\times 10^6\,\Ms$. 
The instrument turns out to
be a rather sensitive galactic monitor; in fact it would be able to
detect secondary black holes with $m_2 \simgt 30\,\Ms$ at SNR $> 5$,
and for $m_2 \simgt 10^3\,\Ms$ with a remarkable SNR $\sim 100$ or
higher (but the detection probability remains low \cite{BVI,Sig}). 
Furthermore,  most of the
systems visible during one experiment would be detectable (independently)
also in one of the other two data sets, as the time to coalescence for the relevant
range of masses and frequencies is $\,\simeq 2995\,
(f_b/5\times 10^{-4}\,{\rm Hz})^{-8/3}\,(m_2/10^2\,\Ms)^{-1}\,
(m_1/2\times 10^6\,\Ms)^{-2/3}\,{\rm days}$.

We can therefore conclude that CASSINI represents a significant improvement
with respect to previous Doppler missions \cite{BBgr14}, and although
can not guarantee a positive detection of GWs, it is definitely
able to provide valuable astrophysical information on massive black holes, which are
unaccessible to any generation of Earth-based detectors.

\end{document}